\documentclass{article}
\usepackage{iclr2026_conference,times}

\makeatletter
\iclrfinalcopy
\makeatother

\usepackage{float}
\usepackage{mathpartir}
\usepackage{bm}
\usepackage{soul}
\usepackage{xspace}
\usepackage{mathtools}
\usepackage{tikz-cd}
\usetikzlibrary{positioning, shapes, arrows, fit}
\usepackage[scaled=0.65]{beramono}
\usepackage{calrsfs}
\usepackage{subdepth}
\usepackage{dashundergaps}
\usepackage[final]{listings}
\usepackage{mleftright}
\usepackage{extarrows}
\usepackage{placeins}
\usepackage[a]{esvect}
\usepackage{colortbl}
\usepackage{tabularx}
\usepackage{multirow}
\usepackage{subcaption}
\usepackage{enumitem}
\usepackage{longtable}
\usepackage{pifont}
\usepackage{cancel}
\usepackage{wrapfig}
\usepackage{fancyvrb}
\usepackage{tcolorbox}
\usepackage{url}

\input{tex-common/proof-helpers}

\newcolumntype{L}[1]{>{\raggedright\let\newline\\\arraybackslash\hspace{0pt}}p{#1}}
\newcolumntype{C}[1]{>{\centering\arraybackslash}p{#1}}

\DeclareSymbolFont{bbsymbol}{U}{bbold}{m}{n}
\DeclareMathSymbol{\bbsemi}{\mathbin}{bbsymbol}{"3B}

\setlist[enumerate]{leftmargin=1.5em}
\setlist[itemize]{leftmargin=1.5em}

\newcommand*{\ttt}[1]{\texttt{#1}}
\newcommand*{\kw}[1]{{\text{\ttt{#1}}}} 

\DeclareTextFontCommand{\textbfit}{%
  \fontseries\bfdefault 
  \itshape
}

\newcommand*{\appref}[1]{Appendix~\ref{app:#1}}

\newcommand*{\figref}[1]{Figure~\ref{fig:#1}}

\newcommand*{\secref}[1]{Section~\ref{sec:#1}}

\newenvironment{nop}{}{}
\newenvironment{sdisplaymath}
   {\begin{nop}\small\begin{displaymath}}
   {\end{displaymath}\end{nop}\ignorespacesafterend}
\newenvironment{smathpar}
   {\begin{nop}\small\begin{mathpar}}
   {\end{mathpar}\end{nop}\ignorespacesafterend}

\newenvironment{mathfig}{\begin{sdisplaymath}}{\end{sdisplaymath}}

\makeatletter
\newbox\sf@box
  {\def\sf@one{#1}%
   \def\sf@two{#2}%
   \setbox\sf@box\hbox
      \bgroup}%
  { \egroup
   \ifx\@empty\sf@two\@empty\relax
     \def\sf@two{\@empty}
   \fi
   \ifx\@empty\sf@one\@empty\relax
      \subfloat[\sf@two]{\box\sf@box}%
   \else
      \subfloat[\sf@one][\sf@two]{\box\sf@box}%
   \fi}
\makeatother

\RequirePackage{xcolor}

\definecolor{highlightcolor}{rgb}{1.0,0.8,0.8}
\definecolor{shadecolor}{rgb}{0.9,0.9,0.9}
\definecolor{lightgray}{rgb}{0.8,0.8,0.8}
\newcommand*{\shadebox}[1]{\fcolorbox{lightgray}{shadecolor}{\raisebox{0pt}[0.60\baselineskip][0.05\baselineskip]{#1}}}

\makeatletter
\newcommand{\superimpose}[2]{%
  {\ooalign{$#1\@firstoftwo#2$\cr\hfil$#1\@secondoftwo#2$\hfil\cr}}}
\makeatother

\newsavebox{\vardisplaymathbox}

\newcommand*{\SuggestionAgent}{SuggestionAgent\xspace}
\newcommand*{\InterpretationAgent}{InterpretationAgent\xspace}
\newcommand*{\SciGen}{SciGen\xspace}
\newcommand*{\gptfour}{\textit{gpt-4o}\xspace}
\newcommand*{\gptfive}{\textit{gpt-5}\xspace}

\newcommand*{\textfrag}[1]{``#1''}

\definecolor{verylightgray}{gray}{0.9}
\definecolor{lightgray}{gray}{0.5}
\definecolor{mediumgray}{gray}{0.45}

\newlength\lsthorizontalpadding
\newcommand*\lstnumberstyle{\ttfamily\scriptsize\textcolor{lightgray}}
\newcommand*{\lbbar}{\{\kern-0.3em|}
\newcommand*{\rbbar}{|\kern-0.3em\}}
\newlength\lstnumbersep
\newlength\lstnumberwidth
\lstdefinelanguage{Fluid}{%
   morekeywords={as,else,fun,if,import,in,let,match,then,error}%
  ,moredelim=[s][\itshape]{`}{`}
}
\lstnewenvironment{codecol}[1][]{\lstset{language=code,#1,moredelim=*[is][\textcolor{mediumgray}]{|}{|}}}{}

\setcitestyle{nosort}

\begin{document}

\title{AI-Assisted Authoring for Transparent, \\Data-Driven Documents}
\author{
   Alfonso Piscitelli \\
   University of Salerno \\
   Fisciano, Italy \\
   \texttt{apiscitelli@unisa.it} \\
   \And
   Joe Bond \\
   University of Bristol \\
   Bristol, UK \\
   \texttt{j.bond@bristol.ac.uk} \\
   \And
   Cristina David \\
   University of Bristol \\
   Bristol, UK \\
   \texttt{cristina.david@bristol.ac.uk} \\
   \And
   Mattia De Rosa \\
   University of Salerno \\
   Fisciano, Italy \\
   \texttt{matderosa@unisa.it} \\
   \And
   Ali Mohammed \\
   Manchester Metropolitan University \\
   Manchester, UK \\
   \texttt{22465707@stu.mmu.ac.uk} \\
   \And
   Federico Nanni \\
   The Alan Turing Institute \\
   London, UK \\
   \texttt{fnanni@turing.ac.uk} \\
   \And
   Jacob Pake \\
   University of Kent \\
   Canterbury, UK \\
   \texttt{jwp24@kent.ac.uk} \\
   \And
   Roly Perera \\
   University of Cambridge \\
   Cambridge, UK \\
   \texttt{roly.perera@cl.cam.ac.uk} \\
   \And
   Jessy Sodimu \\
   University of Bath \\
   Bath, UK \\
   \texttt{js3720@bath.ac.uk} \\
   \And
   Chenyiqiu Zheng \\
   University College London \\
   London, UK \\
   \texttt{chenyiqiu.zheng.23@ucl.ac.uk} \\
}


\maketitle

\begin{abstract}
We introduce \emph{transparent documents}, interactive web-based scholarly articles which allow readers to explore the relationship to the underlying
data by hovering over fragments of text, and present an LLM-based tool for authoring transparent documents, building on recent developments in data
provenance for general-purpose programming languages. As a target platform, our implementation uses Fluid, an open source programming language with a
provenance-tracking runtime. Our agent-based tool supports a human author during the creation of transparent documents, identifying fragments of text
which can be computed from data, such as numerical values selected from records or computed by aggregations like sum and mean, comparatives and
superlatives like ``better than'' and ``largest'', trend-adjectives like ``growing'', and similar quantitative or semi-quantitative phrases, and then
attempts to synthesise a suitable Fluid query over the data which generates the target string. The resulting expression is inserted into the article's
web page, turning the static text fragment into an interactable data-driven element able to reveal the data that underwrites the natural language
claim. We evaluate our approach on a subset of SciGen, an open source dataset consisting of tables from scientific articles and their corresponding
descriptions, which we extend with hand-generated counterfactual test cases to evaluate how well machine-generated expressions generalise. Our results
show that gpt4o is often able to synthesise compound expressions extensionally compatible with our gold solutions.
\end{abstract}

\section{Introduction: Transparent, Data-Driven Documents}

When interpreting or verifying data-driven claims, a key challenge lies in tracing specific claims back to the
relevant data. In peer review, for example, empirical claims typically lack author-supplied links to data,
making them hard for reviewers to check directly~\citep{weber20}. Paper retractions, meanwhile, are often
attributable not to fraud, but to simple errors in data management or analysis~\citep{hu25}. The use of large
language models (LLMs) to interpret scholarly documents has seen considerable attention recently, from
fact-checking~\citep{abu-ahmad25} to interpretation of charts and figures~\citep{roberts24}, but current LLM
interfaces do not support direct interrogation of visual or other outputs for traceability to inputs.

Recent advances in data provenance and data visualisation~\citep{psallidas18smoke,bond25}, on the other hand,
have pushed in this direction using a more infrastructural approach. These approaches link computed outputs
to their data sources directly by tracking dependency information. This allows visual outputs to support
\emph{provenance queries}, user interactions (e.g. mousing over visual elements) that reveal how output
features relate to data. The advantage of this approach is that the relationships to data sources are exposed
automatically via trusted infrastructure, typically a query language or general-purpose programming language
which tracks how data flows through a computation. However, these approaches are limited to outputs computed
from data, such as visualisations. What is missing is a way to extend these ``direct interrogation'' features
to natural language itself, where the main claims of most scholarly articles are actually made.

In this paper, we address this gap by combining two complementary approaches: the ability of LLMs to
understand technical language and synthesise queries over data, plus the provenance-tracking infrastructure of
an open source programming language called Fluid (\url{https://f.luid.org/})~\citep{perera22,bond25}.
Together, these two technologies enable the creation of \emph{transparent documents}, web-based scholarly
articles with two key transparency features:
\begin{enumerate}
\item \textbf{Data-driven:} Quantitative statements expressed in natural language --- e.g.~that system $X$ is
faster than system $Y$ on some task --- are computed from the relevant data, rather than occurring
merely as static fragments of text.

\item \textbf{Data linking:} Readers and reviewers can interactively trace such claims back to the specific
data elements that support them, through embedded provenance queries.
\end{enumerate}

\begin{figure}
    \centering
    \includegraphics[width=\linewidth]{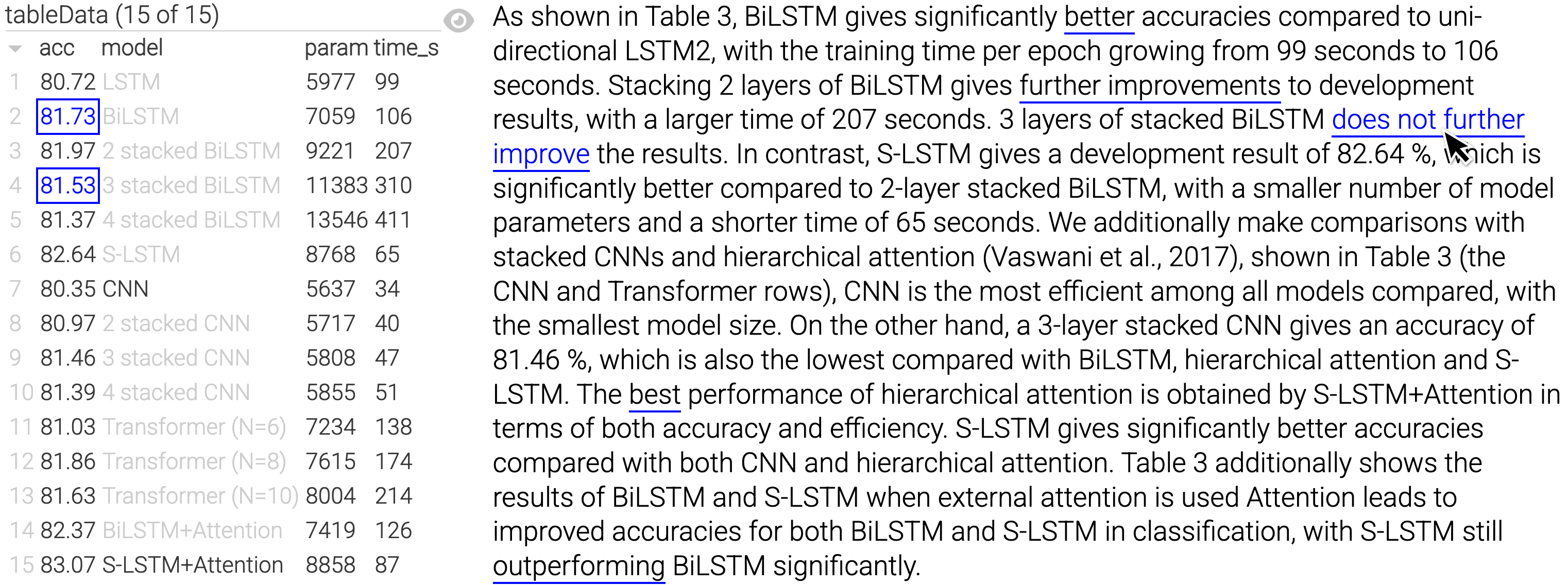}
    \vspace{1mm}
    \hrule
    \includegraphics[width=\linewidth]{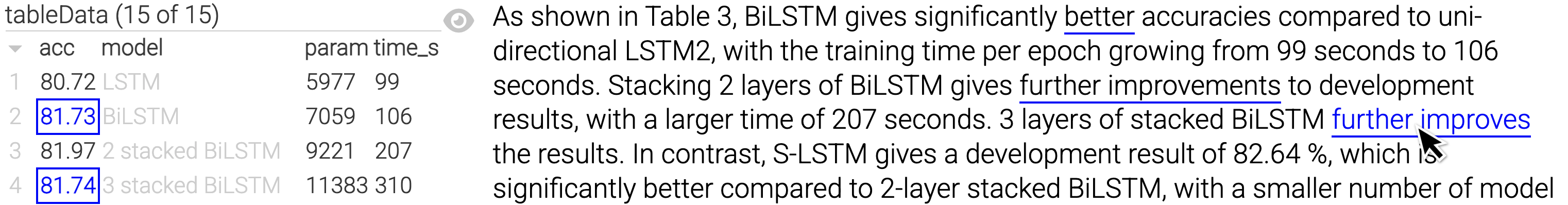}
    \caption{Two versions of a transparent document, showing text fragments linked to data}
    \label{fig:scigen-example-website}
\end{figure}

\figref{scigen-example-website}, generated from our implementation, illustrates these two features. The upper
section shows a ``transparent'' excerpt from \cite{zhang18}, a scholarly article comparing text encoding
techniques. When a reader hovers over the phrase \textfrag{does not further improve}, the relevant data are
highlighted on the left. Other fragments (e.g. \textfrag{better than}, \textfrag{further improvements}) that
refer to the same data are also marked, allowing the reader to explore supporting and contrasting evidence.
The lower section shows a counterfactual situation where the authors' experiments had produced different
results: here the phrase \textfrag{does not further improve} is replaced by \textfrag{further improves}.

\begin{figure}
    \small
    \includegraphics[scale=0.169]{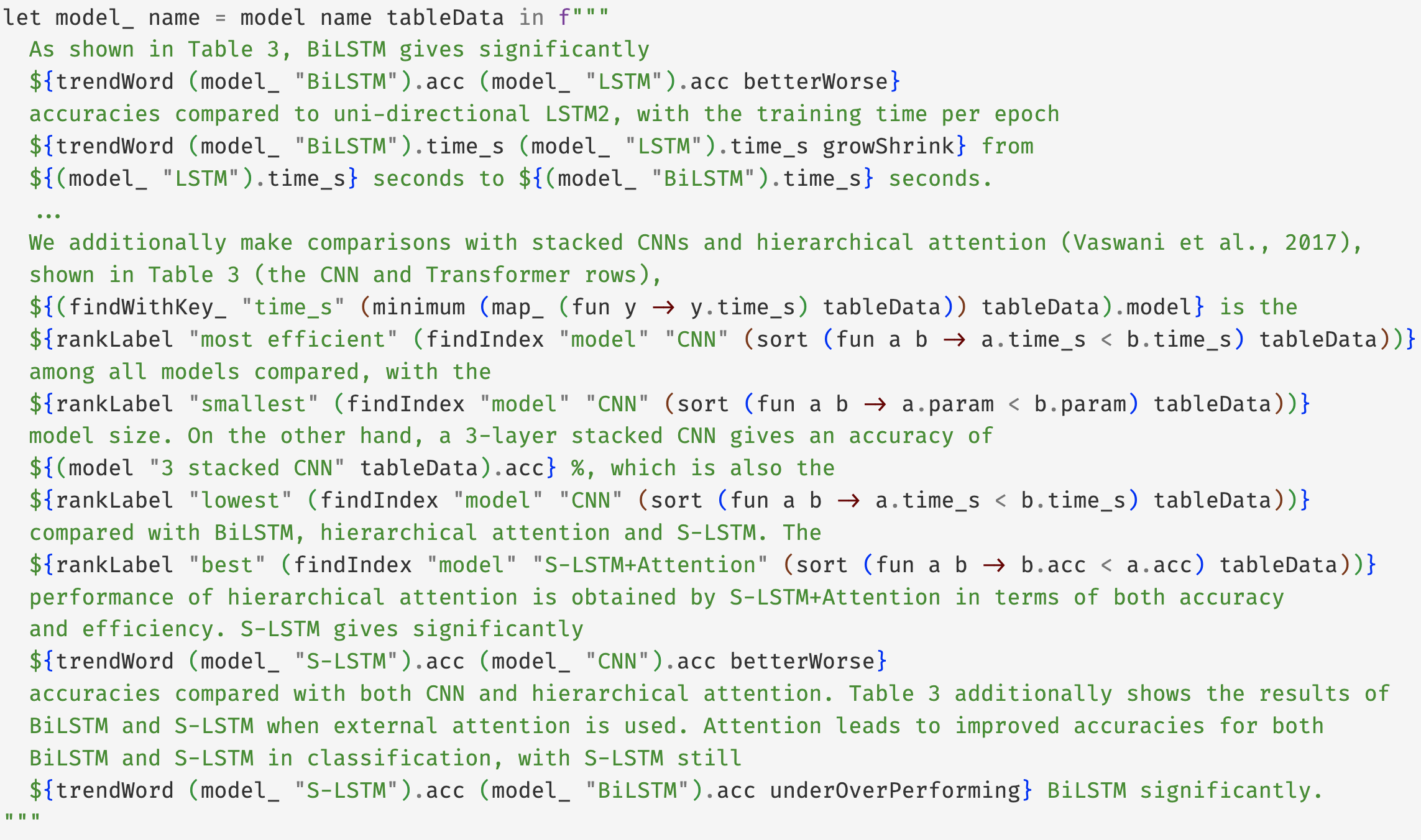}
    \caption{Gold solution for transparent document in \figref{scigen-example-website} (some lines omitted)}
    \label{fig:fluid-example-paragraph}
\end{figure}

This transparent version of the document was implemented in Fluid. The source code is shown in
\figref{fluid-example-paragraph}, and makes use of several helper functions, a representative subset of which
are shown in \figref{fluid-scigen}. What makes our solution interesting is that the provenance-tracking
runtime of Fluid \emph{and} the LLM-based authoring support are both essential components of the solution,
with Fluid providing the interactions, and the LLM-based tool making the authoring process feasible.
Generating code for a traditional language like Python would still result in a data-driven document, but
crucially without the interactive provenance queries; and without AI-based tooling to support the authoring
process, the author would be faced with creating the code in \figref{fluid-example-paragraph} by hand, which
is unlikely to be feasible as part of the usual scientific writing process.

AI-assisted authoring of transparent documents thus support turning static text into interactable, data-driven
content able to expose the evidential basis of scholarly claims. We envisage two use cases. First, when
\textbf{authoring} content for an online article, a journalist or scientific publisher may wish to provide
text which is linked to the underlying data so that the evidence base for the textual claims can be explored
directly from the article. Second, when \textbf{reading} a document reporting on findings derived from open
data (perhaps a scientific paper or climate report), the reader may want to retroactively interpret parts of
the text as queries over the available data and gradually ``rationally reconstruct'' the relationship between
claims in the paper and the evidence base. This might be just to aid their own comprehension, or part of a
formal peer review process.

\paragraph{Contributions.} Our specific contributions are as follows. We leave implementing a full
Copilot-like authoring plugin for an IDE such as VSCode or Cursor for future work (\secref{conclusion}).

\begin{itemize}
\item A proof-of-concept LLM-based tool for iteratively transforming a preexisting opaque document and
associated data set into a transparent, data-driven counterpart (\secref{authoring-workflow});
\item A summary of the natural language idioms we have studied (\secref{nl-idioms}) and an empirical
evaluation of how well state-of-the-art models are able to solve the associated interpretation and code
synthesis problems (\secref{evaluation}).
\end{itemize}

\section{AI-Assisted Authoring Workflow}
\label{sec:authoring-workflow}

\begin{figure}
    \centering
    \includegraphics[width=0.8\linewidth]{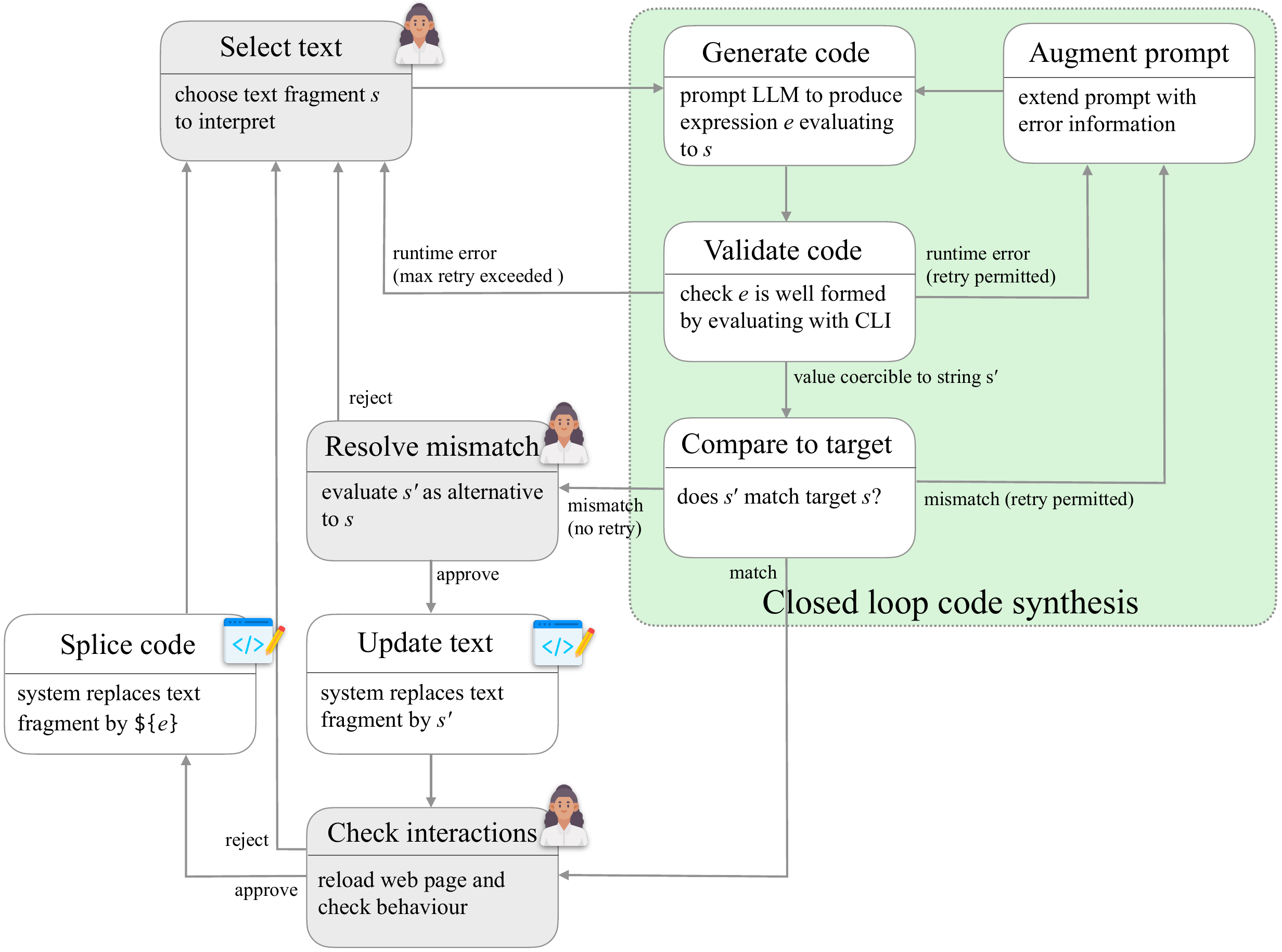}
    \caption{Human-in-the-Loop workflow (states requiring human intervention in grey)}
    \label{fig:interpretation-agent}
\end{figure}

Our authoring tool is composed of two LLM-based agents. A \textbf{\SuggestionAgent} identifies text fragments
potentially computable from data, and an \textbf{\InterpretationAgent}, given a text fragment provided by the
SuggestionAgent or by the author, attempts to synthesise a Fluid expression which computes the target
fragment. The main components of the workflow are as follows:

\begin{enumerate}

\item \textbf{Initial configuration.} The author imports the target text and accompanying data into the system
to create a programmatic representation of the target document. Initially this is simply equivalent to the
target text, taking the form of a string literal \kw{"""..."""}, where the triple quotes are Fluid syntax for
a Python or JavaScript-style \emph{interpolated string}, i.e.~a literal where expressions of the form
\kw{\{$e$\}} are permitted within the string. The \SuggestionAgent analyses the target text and identifies any
fragments which are candidates for being computed instead of remaining as literal substrings.

\item \textbf{High-level Authoring workflow.} The system then enters the human-in-the loop authoring workflow
shown in Figure~\ref{fig:interpretation-agent}, where the author interacts with the \InterpretationAgent. The
system waits for the author to select a fragment of text $s$ to interpret (perhaps previously highlighted by
the \SuggestionAgent). The system then attempts to generate a candidate Fluid expression $e$ using the
closed-loop synthesis step (3) below. If code synthesis succeeds with an expression $e$, the system proceeds
to the manual validation step (4) below. If the synthesis step fails with no expression, no remedial action is
possible; this is considered an unsuccessful path through the workflow and returns the system to the entry
state. Otherwise the synthesis step produces an expression $e$ which evaluates to a mismatched string $s' \neq
s$ outcome, and the user can choose to manually abort and return to the entry state, or optionally to
\emph{revise the goal}, replacing $s$ with $s'$ in the target document and retaining $e$ as the candidate
expression. This is intended to cover the situation where the author has made a claim which is
\emph{incorrect}, and the data set and surrounding natural language have led the LLM to synthesise an
expression which generates a different value from the one specified by the user.

\item \textbf{Code synthesis step.} The expression synthesis step is an error-guided iterative prompting
loop~\citep{skreta23}, beginning with an initial prompt sent to the LLM (see \emph{Prompt design} below)
requesting the generation of an expression $e$. Using the Fluid command-line interface, the expression is
validated to check that it evaluates without error, produces a value coercible to a string $s'$, and finally
that $s'$ is equal to the target fragment $s$. Any failure triggers prompt augmentation with the appropriate
error message and the system retries generation. If code synthesis loop is able to yield an expression which
computes $s$ within a maximum number of retries, the synthesis step succeeds with $e$. If the last generated
$e$ was invalid (resulting in an error), the code synthesis step fails with no expression. Otherwise, code
synthesis produces an expression $e$ but with a mismatched string outcome $s' \neq s$.

\item \textbf{Manual validation step.} Once a candidate expression has been generated, the system replaces the
selected substring $s$ with the interpolation expression \kw{\{$e$\}}, creating a new (but only tentative)
document configuration. The author can republish the web page hosting the document and interact with the
proposed revision. As shown in \secref{evaluation}, this is an important validation step that can reveal
errors in the generated expression. If the interactions look reasonable, the author can approve the new
document state; this is the primary successful path through the workflow and returns the system to the entry
state where it is waiting for another top-level interaction from the author. Otherwise, the author rejects the
proposed change and returns to the entry state without any change to the document.

\end{enumerate}

This human-in-the-loop design combines automated synthesis with validation and author oversight, providing a
substantial level of automation, but requiring the author to intervene at key steps to ensure correctness.

\paragraph{\InterpretationAgent prompt design.}

The \InterpretationAgent is guided by a structured system prompt that frames code generation as a precise
replacement task. The model receives the imported datasets, helper modules, and the current Fluid
representation of the paragraph, in which a text fragment is marked with the tag \kw{[REPLACE …]}. The task is
to substitute this placeholder with a Fluid expression that evaluates exactly to the target string,
reconstructing quantitative or comparative claims as data queries. To ensure integration with the workflow,
the output must consist solely of a syntactically valid Fluid expression, with no additional commentary. The
full prompt is given in~\appref{system-prompt:interpretation-agent}.




\begin{figure}[t]
    \centering
    \begin{subfigure}{0.50\linewidth}
        \centering
        \includegraphics[width=\linewidth]{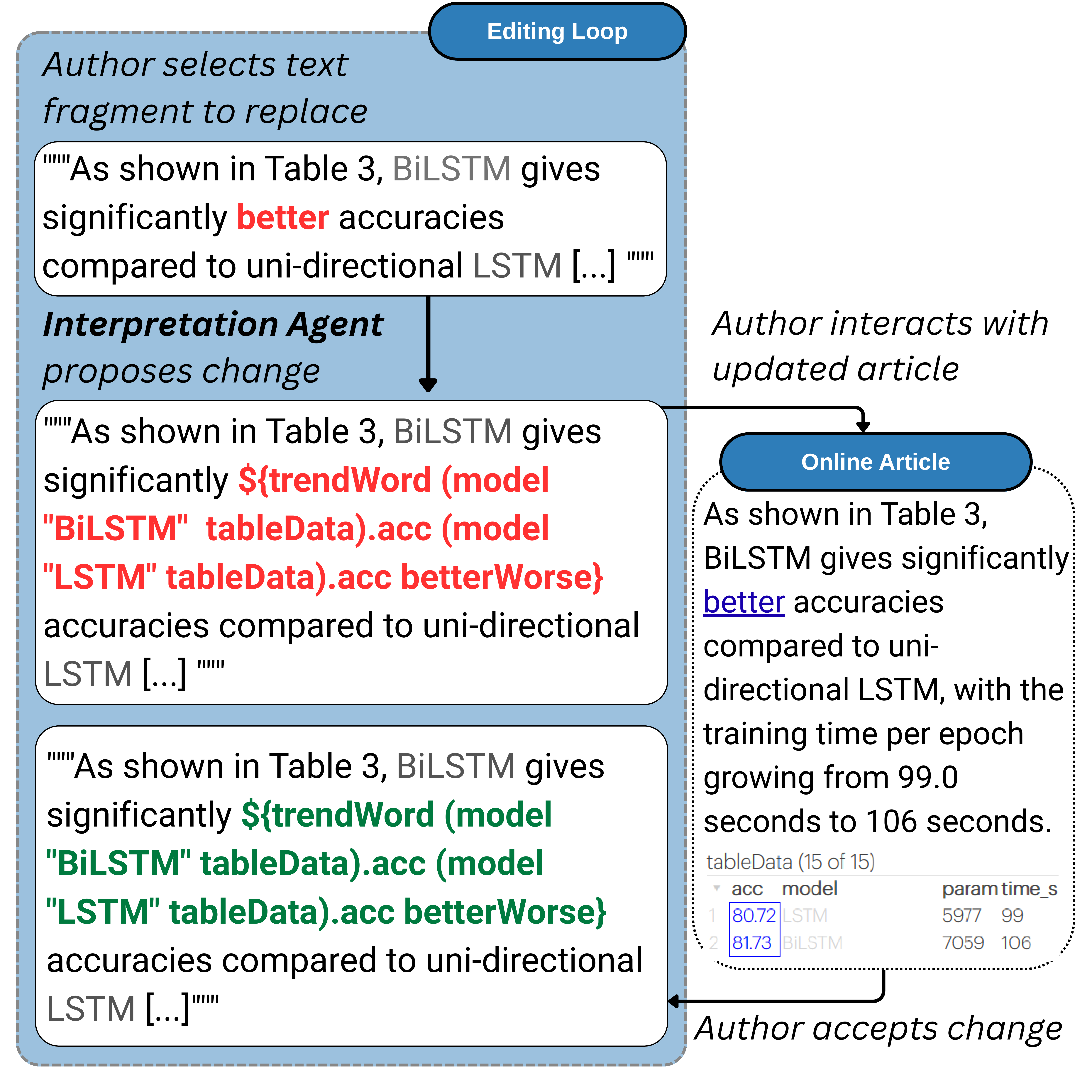}
        \caption{Author accepts expression}
        \label{fig:data-flow-correct}
    \end{subfigure}\hfill
    \begin{subfigure}{0.50\linewidth}
        \centering
        \includegraphics[width=\linewidth]{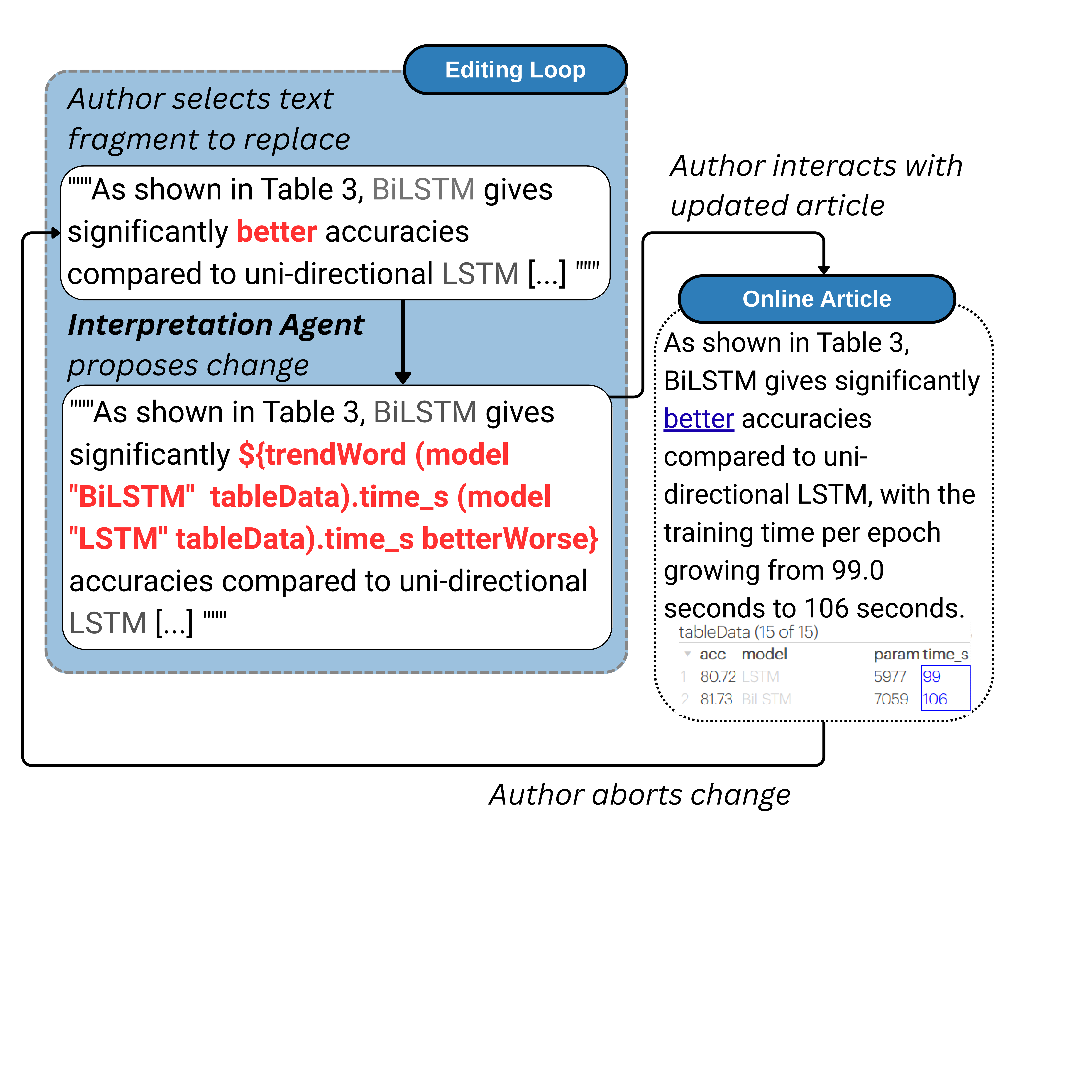}
        \caption{Author identifies error and rejects}
        \label{fig:dataflow-error}
    \end{subfigure}
    \caption{Two possible paths through editing loop, with interactive verification of generated code}
    \label{fig:overall-architecture}
\end{figure}

\section{Target idioms of natural language}
\label{sec:nl-idioms}

\begin{table}[!ht]
    \centering
    \footnotesize
    \renewcommand{\arraystretch}{1.2}
    \begin{tabular}{>{\raggedright\arraybackslash}p{1.4cm} >{\raggedright\arraybackslash}p{5cm} >{\raggedright\arraybackslash}p{6.3cm}}
        \hline
        \textbf{Label}                & \textbf{Example} & \textbf{Gold Solution for Example} \\
        \rowcolor{gray!20}
        Data retrieval
        & the training time per epoch growing from \hl{67} seconds to 106 seconds.
        &
        \begin{lstlisting}[language=Fluid,numbers=none,aboveskip=-7pt,belowskip=-8.5pt]
(model_ "LSTM").time_s
        \end{lstlisting}
        \\
        Ratio &
        The Energy Sector accounts for total methane emissions of \hl{52.80}\% in 2030.
        &
        \begin{lstlisting}[language=Fluid,numbers=none,aboveskip=-7pt,belowskip=-8.5pt]
(getByCategory "Energy Sector" year).emissions /
  sum (map (fun x -> x.emissions)
    (getByYear year)) * 100

        \end{lstlisting}  \\
        \rowcolor{gray!20}
        Average
        & The average methane emissions for the year 2030 is \hl{13.51} &
        \begin{lstlisting}[language=Fluid,numbers=none,aboveskip=-7pt,belowskip=-8.5pt]
sum (map (fun x -> x.emissions)
  (getByYear year)) / length records
        \end{lstlisting} \\
        Min/Max                          & The Energy Sector recorded its highest methane emissions in \hl{2030}             &
        \begin{lstlisting}[language=Fluid,numbers=none,aboveskip=-7pt,belowskip=-8.5pt]
let maxEntry = maximumBy (fun x -> x.emissions)
   (filter (fun x -> x.type == "Energy Sector")
      tableData)
in maxEntry.year
        \end{lstlisting} \\
        \rowcolor{gray!20}
        Rank &
        3-layer stacked CNN gives an accuracy of 81.46\%, which is the \hl{lowest} compared with BiLSTM, and S-LSTM  &
        \begin{lstlisting}[language=Fluid,numbers=none,aboveskip=-7pt,belowskip=-8.5pt]
rankLabel "lowest"
  (findIndex "model" "CNN"
    (sort cmpTime tableData)) \end{lstlisting} \\
        Sum &
        The total methane emissions for the year 2030 is \hl{37.74} for Agriculture &
        \begin{lstlisting}[language=Fluid,numbers=none,aboveskip=-7pt,belowskip=-8.5pt]
sum (map (fun x -> x.emissions)
   (getByYear year))
        \end{lstlisting} \\
        \rowcolor{gray!20}
        Comparison
        & The training time per epoch \hl{growing} from 67 seconds to 106 seconds. &
        \begin{lstlisting}[language=Fluid,numbers=none,aboveskip=-7pt,belowskip=-8.5pt]
trendWord
  (model_ "BiLSTM" tableData).time_s
  (model_ "LSTM" tableData).time_s
  growShrink
        \end{lstlisting} \\
        Generalised quantifiers
        & In the case of one syndrome (Hemorrhagic) we noticed an \hl{unusually low} level of recall for SVM but not for NB. &
        \begin{lstlisting}[language=Fluid,numbers=none,aboveskip=-7pt,belowskip=-8.5pt]
unusuallyHighLow (overallComparison [
  compareCols col "naive_bayes_r"
    (findWithKey_ "synd" "Hem." tableData)
  | col <- ["svm1_r", "svm2_r", "svm3_r", "svmr_r"]
])
        \end{lstlisting} \\
        \hline
    \end{tabular}
    \caption{Quantitative/semi-quantitative natural language forms considered in this paper}
    \label{tab:natural-language-forms}
\end{table}

Table~\ref{tab:natural-language-forms} summarises the natural language idioms studied in this paper. With
state-of-the-art models like \gptfour and \gptfive, our system is able to resolve basic table lookups of direct
numerical values, as well as computations of percentages, averages, minima and maxima, and totals, each mapped
to the corresponding aggregation over the source data. For example, phrases such as \textfrag{the Energy
Sector accounts for 52.80\% of total emissions} and \textfrag{average methane emissions for 2030 is 13.51} are
interpreted in terms of sum and mean respectively over the relevant data values. Similarly, \textfrag{recorded
its highest emissions in 2030} is interpreted as a \kw{maximumBy} query, while a statement such as
\textfrag{CNN gives the lowest accuracy} is mapped to an explicit computation of rank.

We also consider \emph{trend} expressions, which comparative natural language phrases describing how a data
attribute evolves over time, such as \textfrag{training time growing from 67 to 106 seconds}. Such idioms are
mapped to higher-order functions like \kw{trendWord} parameterised on additional helper functions such as
\kw{growShrink} and \kw{betterWorse} (shown in \figref{fluid-scigen}) which map comparisons to appropriate
natural language phrases.

\begin{figure}[t]
\begin{minipage}[t]{0.48\textwidth}
    \small
\begin{lstlisting}
let ordinalMap =
   [ { lastDigit: 1, suffix: "st" },
     { lastDigit: 2, suffix: "nd" },
     { lastDigit: 3, suffix: "rd" } ];

let ordinal n =
   if n <= 0 then error "n <= 0 not supported"
   else if (n < 4) then
      numToStr n ++
      (findWithKey_ "lastDigit" n ordinalMap).suffix
   else if (n >= 4) `and` (n <= 20) then
      numToStr n ++ "th"
   else error "n > 20 not supported";

let rankLabel word n =
   (if n == 1 then "" else ordinal n ++ "-") ++ word;
\end{lstlisting}
\end{minipage}%
\hspace{12mm}%
\begin{minipage}[t]{0.43\textwidth}
\begin{lstlisting}[firstnumber=19]
let trendWord n1 n2 compareWord =
    compareWord (compare n1 n2);

let growShrink EQ = "unchanging";
    growShrink LT = "shrinking";
    growShrink GT = "growing";

let smallerHigher EQ = "equal";
    smallerHigher LT = "smaller";
    smallerHigher GT = "larger";

let improvements EQ = "no further improvements";
    improvements LT = "no further improvements";
    improvements GT = "further improvements";
\end{lstlisting}
\end{minipage}
    \caption{SciGen helper functions (representative examples)}
    \label{fig:fluid-scigen}
\end{figure}

Taken together, these categories cover a representative portion of the numerical reasoning idioms found in the
\SciGen benchmark. However, some linguistic forms that commonly arise in scholarly articles are not covered in
our analysis. We have yet to study approximate quantitative terms like \textfrag{around 50\%} or
\textfrag{roughly 100 instances}, nor interval-based descriptions such as \textfrag{between 30 and 40\%} or
\textfrag{within 5–10 seconds}. While we have no reason for thinking these will present specific difficulties,
other forms are likely to be more challenging. So-called \emph{graded} modal adverbs~\citep{lassiter17} which
modify adjectival comparatives like \textfrag{better} -- as in \textfrag{slightly better} and
\textfrag{significantly higher} -- especially when combined with trends over time, as in \textfrag{steadily
increasing} or \textfrag{sharply declining} -- are likely to prove difficult because the interpretation of
these qualifiers can be subjective and context-dependent. Generalised quantifiers like \textfrag{generally}
and \textfrag{usually}~\citep{barwise81} present similar challenges because colloquial use may differ from
more formal uses (in some situations ``most'' might mean a majority, i.e.~greater than 50\% of cases, but in
others may mean only ``greater than any other alternative proportion''). On the other hand these difficulties
also present themselves to human readers, so extending coverage to these idioms would substantially deepen our
tool's ability to bridge natural language reporting with interpretation in terms of the underlying dataset,
perhaps revealing inconsistent use of technical language on the part of the author. We discuss this further in
\secref{conclusion}.

\section{Experimental Evaluation}
\label{sec:evaluation}

\subsection{Research questions}

Our evaluation tests the ability of the \InterpretationAgent to translate quantitative and semi-quantitative
expressions from scholarly natural language into executable queries that operate on the underlying dataset.
Beyond raw accuracy, we are also concerned with how performance varies with task complexity, and whether the
generated expressions are robust under changes to data or in the presence of ambiguity or other low data
quality issues. These are captured in two research questions:

\paragraph{RQ1. Interpretation Accuracy across Linguistic Idioms and Complexity.}

To what extent can LLMs accurately interpret quantitative and semi-quantitative claims in scholarly text as
data queries? We examine performance across a range of linguistic idioms (e.g. averages, percentages, min/max,
ranks, as summarised in Table~\ref{tab:natural-language-forms}) and investigate how accuracy varies with task
complexity, measured (somewhat crudely) by the number of query sub-expressions (e.g. retrieval, aggregation,
or arithmetic) present in the gold solution.

\paragraph{RQ2. Generalisability and Robustness.}

How well do the generated expressions generalise when the underlying data changes, or when the input contains
misleading or ill-specified information? We test whether generated queries continue to produce correct outputs
under a set of hand-generated counterfactual modifications of the dataset, based on expected query results
specific to each test case, and also how counterfactual performance is impacted by the presence of misleading
or adversarial phrasing. Table~\ref{tab:problematic_cases} shows some of cases we deem problematic in this
sense; in these cases, producing a valid expression is likely to be challenging because of ambiguities in the
input data or accompanying natural language.

\begin{figure}
    \centering
    \begin{subfigure}{0.48\linewidth}
        \centering
        \includegraphics[width=\linewidth]{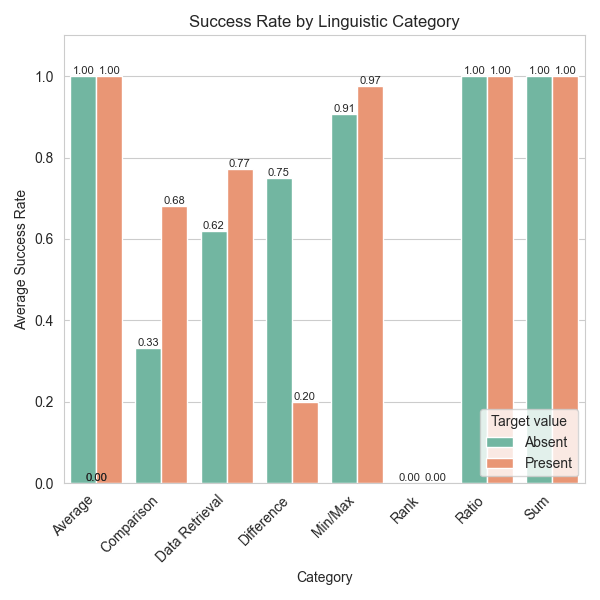}
        \caption{Success rate by Linguistic Category}
        \label{fig:success_rate_by_category}
    \end{subfigure} \hfill
    \begin{subfigure}{0.48\linewidth}
        \centering
        \includegraphics[width=\linewidth]{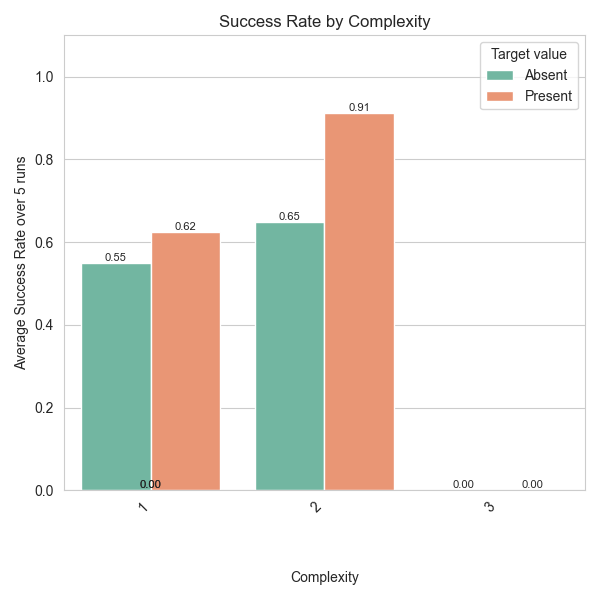}
        \caption{Success rate by Complexity}
        \label{fig:success_rate_by_complexity}
    \end{subfigure}
    \caption{Success rate of the proposed system, measured over 5 runs with \gptfour}
    \label{fig:success_rate_comparison}
\end{figure}

\subsection{Results}

\paragraph{Interpretation Accuracy across Linguistic Idioms and Complexity.} To evaluate RQ1, we used a sample
of the \SciGen dataset~\citep{moosavi21}, an open source dataset consisting of tables from scientific articles
and their corresponding descriptions. We aggregated the results according to the linguistic categories from
Table~\ref{tab:natural-language-forms}. Figure~\ref{fig:success_rate_by_category} illustrates the success rate
for each category, both with and without target-value sharing.

The results show that \emph{the system is robust when provided with sufficient guidance but degrades when
underspecified.} With the target-value sharing, the InterpretationAgent produced correct Fluid expressions in
74.9\%  (S.D. 3.0\%) of cases, but performance dropped to 57.1\% when the target was withheld. This highlights
the system's reliance on explicit cues when resolving ambiguous fragments.

\emph{Performance also varied across linguistic categories.} Success rates exceeded 68\% for comparison,
77.3\% for data retrieval, and 97\% for min/max search tasks. In contrast, accuracy decreased significantly
for expressions requiring differences (20\%) and for ranking tasks (0\%).

The trend for compositional complexity is more nuanced as shown in
Figure~\ref{fig:success_rate_by_complexity}, which reports the success rate as a function of the number of
categories assigned to each expression: success rates are 62\% for single-category expressions, increase to
91\% when two categories are combined, but collapse to 0\% when three categories are involved. This suggests
that \emph{moderate composition can actually aid performance, perhaps by giving the model clearer structural
cues, but that complexity beyond a certain threshold overwhelms the synthesis process}.




\paragraph{Generalisability and Robustness.} As a preliminary attempt to address RQ2, we carried out
\emph{counterfactual testing} to evaluate the robustness of generated expressions under changes to the
underlying data. In this setup, the input tables were modified according to hand-craft test specifications,
and both the expected and generated expressions were re-executed to check whether the behaviours remained
consistent. Across 300 test executions, 121 contained at least one counterfactual error (an average of 3.8 per
case), of which 42 ultimately still succeeded. These tests highlight cases where an expression may
coincidentally yield the correct output on the original data but fails to be extensionally equivalent more
generally (i.e.~under perturbation). For example, in one test the system generated

\hspace{3mm} \kw{(findWithKey\_ "model" "LSTM2" tableData).time\_s}

intended to retrieve the execution time of the \texttt{LSTM} model, but incorrectly referred to \texttt{LSTM2}.
Counterfactual testing exposed this mismatch, which would otherwise have gone undetected.

At present, counterfactual tests are used only as an evaluation device, not as part of the authoring workflow
itself. For future work (\secref{conclusion}), we plan to investigate automatic generation of counterfactual
tests, allowing these additional robustness checks to be integrated into the document authoring workflow.

\begin{table}[t]
    \centering
    \small
    \begin{tabular}{l p{6cm} p{4cm}}
        \hline
        \textbf{Problem Type} & \textbf{Example} & \textbf{Explanation} \\
        \hline
        false comparison &
        BiLSTM is \hl{the most efficient} among all models compared, with the highest model size &
        BiLSTM is not the most efficient, nor does it have the largest size. \\

        wrong numerical value &
        LSTM is the fastest model with overall time taken being \hl{90} seconds &
        It is not 90 but 106. \\

        ambiguous referent &
        LSTM is the fastest model with overall time taken being \hl{90} seconds &
        There are two type of time in the dataset (training\_time, execution\_time), both with a value of 90 seconds. \\

        \hline
    \end{tabular}
    \caption{Categories of problematic example}
    \label{tab:problematic_cases}
\end{table}

\section{Related Work}
\label{sec:related}

\paragraph{Argument mining.}
\emph{Argument mining} is an area of NLP which involves identifying argumentative structures in text, such as
claims, premises, and conclusions, and mapping them to formal
representations~\citep{palau2009argumentation,lippi2015argument}. Early work focused on rule-based approaches,
while more recent work has leveraged machine learning and deep learning
techniques~\citep{stab2014identifying,stab2017parsing,eger2017neural}. The field has also emphasized defining
annotation schemes for the task, such as the Argumentative Zoning
framework~\citep{teufel2002summarizing,teufel2009towards}, as well as schemes more directly tailored to
argument mining~\citep{stab2014annotating}. The field has focused on various domains, starting from legal
texts~\citep{toulmin2003uses}, and has relied on online resources such as
Debatepedia~\citep{cabrio2013natural}. The community has also rapidly engaged with work that explores the use
of argument mining in scientific texts~\citep{liakata12,lauscher2018investigating} to better understand the
structure of scientific arguments and the relationships between different claims and evidence. While the
advent of LLMs has improved performance~\citep{gorur2025can,vrakatseli2025can}, argument mining remains a
challenging task, particularly when it comes to identifying implicit argumentative relations between discourse
units, and reasoning about relationships among different argumentative components, especially in cross-domain
settings where models struggle to generalise~\citep{gemechu2024aries}. While in our work we do not directly
perform traditional argument mining, this work similarly relies on the identification of claims in text, which
we evaluate following established practices in the field.

\paragraph{NLP and scientific writing.}
The intersection of NLP and scientific writing has gained increasing attention in the last decade, with a
focus on improving the clarity, coherence, and overall quality of scientific texts. On the authoring side,
tools such as automated writing assistants can support researchers in producing more fluent and accessible
text, for instance through grammar correction, summarisation and text
simplification~\citep{napoles2017jfleg,stiennon2020learning,takeshita2024cross,saggion2017automatic}. Other
approaches specifically target the argumentative structure of scientific papers, helping writers to organise
contributions and claims more effectively~\citep{lauscher2018arguminsci}. Nowadays, general purpose LLMs such
as ChatGPT or tools tailored for the task such as Grammarly show the variety of support that NLP tools can
provide to authors~\citep{wu2023chatgpt,ahn2024transformative,khalifa2024using}.

At the same time, NLP methods are being developed to assist reviewers and editors in evaluating submissions.
These include systems for detecting potential issues such as lack of clarity, weak argumentative support, or
even factual inconsistencies and up to scientific fraud~\citep{thakkar2025can, fromm2021argument,
freedman2024detecting}. Such tools can also facilitate meta-reviewing by providing summaries of peer reviews
and identifying points of disagreement among reviewers~\citep{kumar2023reviewers}. While AI tools show
promises in improving the peer-review process~\citep{tyser2024ai}, there are also various risks associated
such as breaches of confidentiality, lack of transparency and biases~\citep{perlis2025artificial}. Our current
work situates itself in the context of NLP tools for supporting the understanding of scientific writing;
specifically, it addresses one of the major critiques toward the automation of such process by offering a
transparent way of examining its workflow.

\paragraph{Interpretable NLP.}

As \figref{scigen-example-website} illustrates, scientific texts routinely make use of comparatives like
``faster'' while leaving one of the argument slots implicit, with the context determining the omitted
referent. LLMs demonstrate considerable competence in resolving these and other more syntactic forms of
anaphora such as pronouns~\citep{zhu25}, but the resolved referent itself -- concretely, what was being
referred to -- remains implicit. Interpretable NLP is a recent research direction which aims to support
comprehension (and production) of text in a more explicit and transparent way~\citep{yulan23}. By generating
code that formalises the interpretation of a comparative like ``faster'', our approach also makes these
implicit references explicit; combining our system with interpretable NLP would allow the user to explore the
linguistic interpretation as well.

\section{Conclusions and Future Work}
\label{sec:conclusion}

We introduced a proof-of-concept system for authoring transparent, data-driven documents by combining LLM-based code synthesis with Fluid's provenance-tracking runtime. Our evaluation on SciGen shows that the approach can reliably link natural language claims to their underlying data, while also revealing common failure modes such as ambiguity and misleading input.


Future work includes reducing reliance on predefined helper functions such as \kw{growShrink} and
\kw{trendWord}. While there is an advantage in using a predefined set of helpers (in that they offer a uniform
framework for interpreting a given scholarly document), we also aim to enable the system to operate in their
absence, for instance by turning ``definition not found'' errors into augmented prompts that trigger automatic
generation of missing definitions. We also plan to broaden the scope of supported artifacts, extending
interpretation to visualisations and intermediate datasets derived from cleansing or aggregation, and to cover
additional idioms such as cardinals, multiplicatives, rounding, and graded adjectives.

Another priority is improving integration and validation. Embedding
the system into developer and authoring environments such as VSCode or
Cursor would make the workflow more seamless, while automatic
generation of counterfactual test cases could strengthen validation at
authoring time. Finally, distinguishing between \emph{referential terms} with
fixed denotations and queries with data-dependent values may help in
repairing false or inconsistent statements, ensuring that generated
expressions remain aligned with both the data and the author's intent.

\section*{Reproducibility Statement.}
To facilitate reproducibility, we provide a zip archive in the supplementary materials containing the complete source code,
the datasets used in our experiments, and a README file with detailed instructions for running the scripts.

\clearpage
\bibliography{tex-common/bib}

\clearpage
\appendix
\section*{Appendices}
\section{\InterpretationAgent System Prompt}
\label{app:system-prompt:interpretation-agent}

\begin{Verbatim}[fontsize=\small]
You are a specialized language model for the Fluid functional programming language.
Your task is to analyze a JSON object that represents the user’s Fluid program and its context,
and to generate the Fluid expression that must replace the [REPLACE value=] placeholder inside
the paragraph.

Input Structure
The JSON input always contains:
-datasets: one or more JSON-like arrays containing the data used by the program
(scenario-related key–value pairs).
-imports: Fluid helper libraries provided by the user’s program.
-code: Additional Fluid functions and definitions from the user’s program.
-paragraph: A description that includes exactly one [REPLACE ...] tag.
-paragraphValue: The correct final version of the paragraph (ground truth).

Note: imports, code, and datasets are part of the user’s Fluid program, not just supporting context.
Your output must be consistent with these definitions.

Task
Identify the [REPLACE ...] tag in paragraph.
If the tag has the value property, generate a Fluid expression that evaluates exactly to that value.
If not, infer the correct value by comparing paragraph, paragraphValue, and (if needed) datasets.
The result must always be a Fluid expression that evaluates to a string.

Output Format
Return only the Fluid expression, nothing else.

Constraints
-Output exactly one valid Fluid expression.
-Ensure it is syntactically correct and consistent with the provided imports and code.
\end{Verbatim}

\section{\SuggestionAgent System Prompt}

\begin{Verbatim}[fontsize=\small]
You are an expression detector for Fluid language.
Fluid is a functional programming language used to represent structured data queries and comparisons in a
transparent way.

TASK DESCRIPTION

Given a natural language paragraph and a structured dataset, identify and annotate the parts of the
paragraph that can be replaced by a Fluid expression.

You must detect:
- Explicit values (e.g., scores, names, numbers)
- Comparative expressions (e.g., *better than*, *worse*, *higher*, *more than*)
- Superlative or aggregated expressions (e.g., *the best*, *highest*, *maximum*, *top performer*)

FORMAT

Replace each detected expression with:

[REPLACE value=...]

Where `value` contains the **original text** of the expression (e.g., "91.57", "better", "the best") —
not the rewritten logic or Fluid code.

IMPORTANT RULE

When replacing comparative or superlative expressions (like "better", "worse", "the best", "highest"),
the `value` **must be the exact original word or phrase** from the paragraph.

Correct:
S-LSTM gives [REPLACE value="the best"] reported results.
BiLSTM performs [REPLACE value="better"] than LSTM.

Incorrect:
S-LSTM gives [REPLACE value="getMaxBy f1 data"] results.
BiLSTM performs [REPLACE value="BiLSTM.acc > LSTM.acc"] than LSTM.

If needed, annotate separate values independently:

Example:
BiLSTM gives [REPLACE value="91.2"]% accuracy, which is [REPLACE value="better"] than LSTM.

---

EXAMPLES

Example Fluid code:

let bestModel = getMaxBy f1 data in bestModel.model

---

INPUT EXAMPLE

Paragraph:
For NER (Table 7), S-LSTM gives an F1-score of 91.57% on the CoNLL test set, which is significantly
better compared with BiLSTMs. Stacking more layers of BiLSTMs leads to slightly better F1-scores
compared with a single-layer BiLSTM. Our BiLSTM results are comparable to the results reported
by Ma and Hovy (2016) and Lample et al. (2016).
In contrast, S-LSTM gives the best reported results under the same settings.
In the second section of Table 7,Yang et al. (2017) obtain an Fscore of 91.26%.

Data:
[
  {model: "BiLSTM", f1: 90.96},
  {model: "2 stacked BiLSTM", f1: 91.02},
  {model: "3 stacked BiLSTM", f1: 91.06},
  {model: "S-LSTM", f1: 91.57},
  {model: "yang2017transfer", f1: 91.26}
]

---

OUTPUT EXAMPLE

For NER (Table 7), S-LSTM gives an F1-score of [REPLACE value=91.57]% on the CoNLL test set,
which is [REPLACE value="better"] compared with BiLSTMs.
Stacking more layers of BiLSTMs leads to [REPLACE value="better"] F1-scores compared with a single-layer BiLSTM.
Our BiLSTM results are comparable to the results reported by Ma and Hovy (2016) and Lample et al. (2016).
In contrast, S-LSTM gives [REPLACE value="the best"] reported results under the same settings.
In the second section of Table 7, Yang et al. (2017) obtain an Fscore of [REPLACE value=91.26]%.
\end{Verbatim}

\end{document}